\documentclass[aps,pre,reprint]{revtex4-1}

\usepackage[utf8]{inputenc}

\usepackage{graphicx}
\usepackage{amssymb,amsmath}
\usepackage{textcomp}
\usepackage[breaklinks]{hyperref}
\usepackage{breakurl}

\usepackage[usenames,dvipsnames]{color}

\begin{document}
\author{Thomas A Caswell}
\affiliation{The James Franck Institute and Department of Physics,\\ The University of Chicago, Chicago, Illinois 60637, USA}

\title{Dynamics of the Vapor Layer Below a Leidenfrost Drop}
\date{\today}

\begin{abstract}

  In the Leidenfrost effect a small drop of fluid is levitated above a sufficiently hot surface, on a persistent vapor layer generated by evaporation from the drop.  The vapor layer thermally insulates the drop from the surface leading to extraordinarily long drop lifetimes.  The top-view shape of the levitated drops can exhibit persistent star-like vibrations.  I extend recent work [Burton et al. PRL 2012] to study the bottom surface of the drop using interference-imaging.  In this work I use a high-speed camera and automated image analysis to image, locate and classify the interference fringes.  From the interference fringes I reconstruct the shape and height profile of the rim where the drop is closest to the surface.  I measure the drop-size dependence of the planar vibrational mode frequencies, which agree well with previous work.  I observe a distinct breathing mode in the average radius of the drop, the frequency of which scales differently with drop size than the other modes.  This breathing mode can be tightly coupled to a vertical motion of the drop.  I further observe a qualitative difference in the structure and dynamics of the vertical profile of the rim between large and small drops.

\end{abstract}

\maketitle

\section{Introduction}
\begin{figure}
  \includegraphics{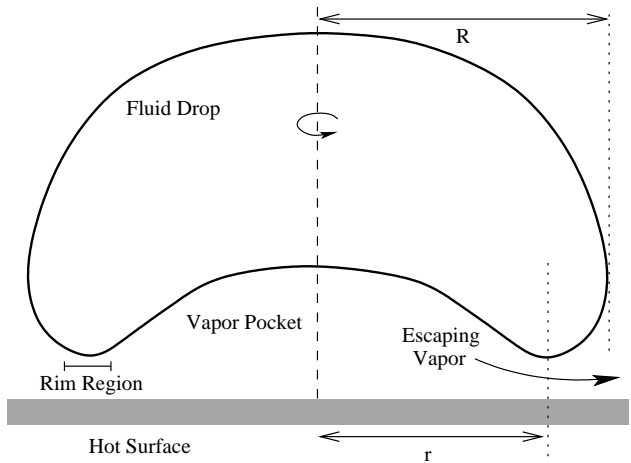}
  \caption{Schematic cross section of an axi-symmetric Leidenfrost drop.  The drop is levitated on a high-pressure vapor pocket which is feed by evaporation from the drop and drained through the narrow gap between the drop and the surface at the rim.  The pressure is maintained due to the resistance to vapor flow through the gap.  The radius of the drop, as observed from above, is denoted as $R$ and the radius from the center of the drop to the rim, as observed from below, is denoted as $r$.  I observe interference between the top of the prism and the bottom of the drop in the rim region where the drop surface and prism surface are nearly parallel}
    \label{fig:drop_scheme}
\end{figure}

A fluid drop placed on a heated surface will rapidly evaporate.  The drop lifetime will decrease with increasing surface temperature until the temperature reaches the Leidenfrost temperature, at which point the drop lifetime increases dramatically~\cite{walker13:_fundam_physic}.  Above the Leidenfrost temperature a persistent vapor layer forms between the drop and the surface; this levitates and insulates the drop.  To support the drop against gravity, the pressure in the vapor layer must be above the ambient pressure. This over-pressure  deforms the bottom surface of the drop into an inverted bowl, shown schematically in Fig.~\ref{fig:drop_scheme}.  The high-pressure region persists due to the resistance to vapor flow through the narrow gap between the drop and the surface, marked as the rim region in Fig.~\ref{fig:drop_scheme}.  The dynamics of the shape and the vertical profile around this rim were measured using high-speed interference-imaging.

Leidenfrost drops are of considerable interest from applied as well as fundamental points of view.  In any application where a working fluid is used to cool a surface, either by spraying or by immersion~\cite{Vakarelski2012}, once the surface reaches the Leidenfrost temperature the heat-transfer, and hence the cooling, is suppressed.  This  can have catastrophic consequences as the liquid becomes less able to transfer heat away from the surface. Because the dynamics of gap between the hot surface and the drop is dominated by the dynamics of the vapor layer, Leidenfrost drops provide an accessible system to study such thin-film vapor dynamics.

The average shape and gap height of an axi-symmetric Leidenfrost drop are set by the combination of the surface tension, $\gamma$, acceleration due to gravity, $g$, the fluid density, $\rho$, and the pressure under the drop.  There are two shape regimes determined by when the drop radius is above or below the capillary length, $\lambda_{c} = \sqrt{\gamma/\rho g}$.  For drops with a characteristic radius smaller than $\lambda_{c}$ surface tension dominates the shape and the drop is spherical, with only a small dent in the bottom.  For drops larger than  $\lambda_{c}$, gravity dominates and the drop flattens into a pancake of height $2\lambda_{c}$ and the radius set by the drop volume~\cite{biance03:_leiden}.  There is a maximum drop size, above which the vapor layer rises through the center of the liquid turning the drop into a torus.  In both regimes, by balancing the evaporative flux into the vapor layer with the flux out due to the pressure difference, the average height of the drop above the surface has been predicted in good agreement with experiment~\cite{biance03:_leiden,  celestini12:_take_off_small_leiden_dropl}.

The static drop can be divided into an inner and outer region separated by the rim, where the surface of the drop is parallel to the hot surface.  The shape of the outer region is that of a non-wetting drop, where the shape can be predicted from the Young-Laplace equation~\cite{burton12:_geomet_vapor_layer_under_leiden_drop, snoeijer09:_maxim}.  The shape inside the rim is set by the pressure in the vapor layer and the surface tension.  By matching these two solutions the full shape of drops near the maximum size can be predicted~\cite{snoeijer09:_maxim} in good agreement with measurements~\cite{burton12:_geomet_vapor_layer_under_leiden_drop,biance03:_leiden}.

Recent work has shown that the height profile around the rim is not axi-symmetric~\cite{burton12:_geomet_vapor_layer_under_leiden_drop,C2SM25656H}.  In this paper I extended this work to study the dynamics of the rim shape and profile.  I have developed image-processing techniques to automate the analysis of the high-speed interference images. These techniques are used to measure the frequency of the shape oscillations of the rim as a function of drop size.  The results are in good agreement with previous predictions and measurements taken from the outer radius of the drop~\cite{holter52:_vibrat_evapor_liquid_drops,tokugawa94:_mechan_self_induc_vibrat_liquid,takaki85:_vibrat_flatt_drop,adachi84:_vibrat_flatt_drop,bouwhuis13:_oscil}. A breathing mode oscillation in the average radius of the rim is also observed.  Its frequency scales differently with drop size than the other modes.  This breathing mode can be strongly coupled to the drop's vertical motion.

The height profile around the rim evolves more slowly than the shape of the rim and than the expected period for capillary waves of comparable size.  Further, there is a qualitative change in the structure and dynamics of the rim as a function of drop size.  Large drops have `active' profiles, with many parts of the rim moving vertically, whereas small drops have more quiescent profiles, with a dominate `frozen-in' low-spatial-frequency fluctuation.

Section~\ref{sec:methods} describes the experimental apparatus and the image-analysis methods.  Section~\ref{sec:results} reports the vibration-mode frequencies and then presents the results on the breathing mode of the rim and the structure and dynamics of the rim profile.

\section{Methods}
\label{sec:methods}
\subsection{Experimental Setup}

\begin{figure*}
  \includegraphics{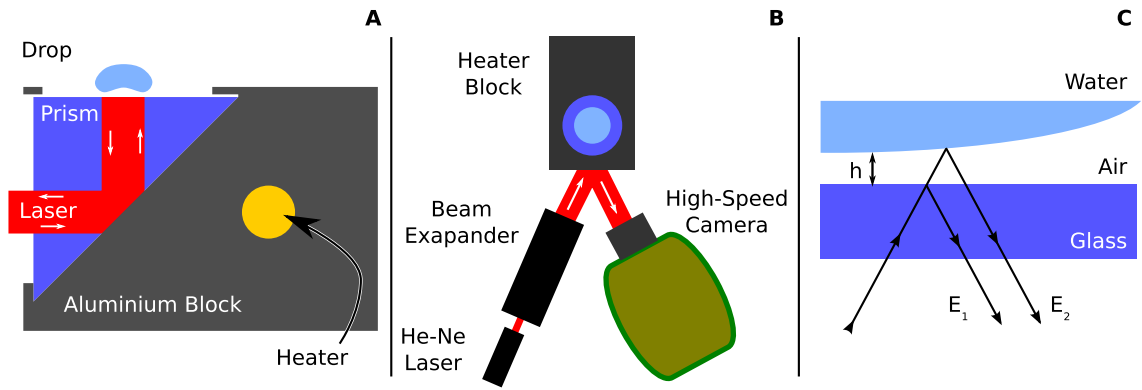}
  \caption{Schematic of experimantal apparatus. Panel A shows a cut-away side view of the heater block.  The incoming laser is reflected off of the hypotenuse of the prism towards the bottom of the drop and the reflections return along the same beam path, out towards the camera.  Panel B shows a plan-view of the experiment. The laser is at an angle to avoid interference between the front and top faces of the prism.  Panel C shows a schematic of light interference in a narrow air-gap.}
  \label{fig:scheme}
\end{figure*}

We use high-speed interference imaging~\cite{driscoll11:_ultraf_inter_imagin_air_splas_dynam} to study the fluctuations of the gap between the bottom of the Leidenfrost drop and the surface. Following~\cite{burton12:_geomet_vapor_layer_under_leiden_drop}, I use a glass prism, heated by an aluminum block with an embedded resistive heater, as the hot surface, as shown in Fig.~\ref{fig:scheme}A.  The drop is illuminated from below with a He-Ne laser~($\lambda_{l} = 632\rm{nm}$) which is reflected up towards the drop off the hypotenuse of the prism.  The laser beam comes in at an angle of approximately 10-15\textdegree~with respect to the normal of the front face of the prism in the horizontal plane, Fig.~\ref{fig:scheme}B, in order to avoid interference between the prism faces.  The reflections from the top surface of the prism and the bottom surface of the drop interfere with each other and produce fringes that can be directly related to variation in the thickness of the vapor layer.   Observation of heavily dyed drops verified that the reflection from the bottom surface dominates the interference pattern.  These interference fringes are directed onto a high-speed camera and imaged at a rate between 1,000 and 10,000 frames per second for 3 to 30 seconds.

In addition to thermally insulating the drop, the vapor layer essentially eliminates friction between the drop and the surface.  Thus any infinitesimal tilt in the prism results in motion of the drop.  I took data either by filming the drop as it transited the field of view or by pinning the drop either by letting it rest against a wall by gravity or by a piece of wire from above.  While these methods have different boundary conditions, the drops are qualitatively the same in all cases.

There are two natural radii in a Leidenfrost drop: the radius as viewed from the top, $R$, marked in Fig.~\ref{fig:drop_scheme}, and the distance from the center of the drop to the rim, $r$, also marked in Fig.~\ref{fig:drop_scheme}.  Most previous work viewed the drop from above where $R$ can easily be measured, but in that configuration the rim on the bottom surface is not visible.  Using interference imaging $r$ can be measured.  As discussed above, the outer profile of a static Leidenfrost drop is the profile of a sessile drop with a contact angle of 180\textdegree.  The numerical relationship between $r$ and $R$ has been established from computed drop profiles~\cite{snoeijer09:_maxim,burton12:_geomet_vapor_layer_under_leiden_drop}. This relationship is used to convert the measurements of $r$ to $R$ in order to compare with previous experiments and theory.

Figure~\ref{fig:scheme}C shows a schematic of optical interference due to a thin air gap of height $h$.  The light reflected off the glass-air interface, $E_{1} \propto A_{1}e^{\imath\psi}$, interferes with the light reflected off of the air-water interface, $E_{2} \propto A_{2}e^{\imath(\psi +\pi + (2\pi/\lambda_{l}) 2h )}$.  The additional phase in $E_{2}$ is from the reflection at the air-water interface ($\pi$) and the phase accumulated from the additional distance the light travels in traversing the air gap twice: ($(2\pi/\lambda_{l}) 2h$).  The  observed intensity is thus $I = |E_{1} + E_{2}|^{2}$, which reduces to
\begin{equation}
  \label{eq:interference}
  I = A_{1}^{2} + A_{2}^{2} - 2A_{1}A_{2}\cos\left( \pi\frac{4h}{\lambda_{l}} \right).
\end{equation}
In the case $A_{1} \approx A_{2}$ and the expression further reduces to
\begin{equation}
  \label{eq:inter_reduced}
  I = 2 A_{1}^{2}\left(1 - \cos\left( \pi\frac{4h}{\lambda_{l}} \right)\right).
\end{equation}
The interference oscillates from light to dark when $h$ varies by $\lambda_{l}/4$.  To normalize the interference images, each pixel is divided by the average for that pixel across all of the frames of a given movie.

\begin{figure}
  \includegraphics{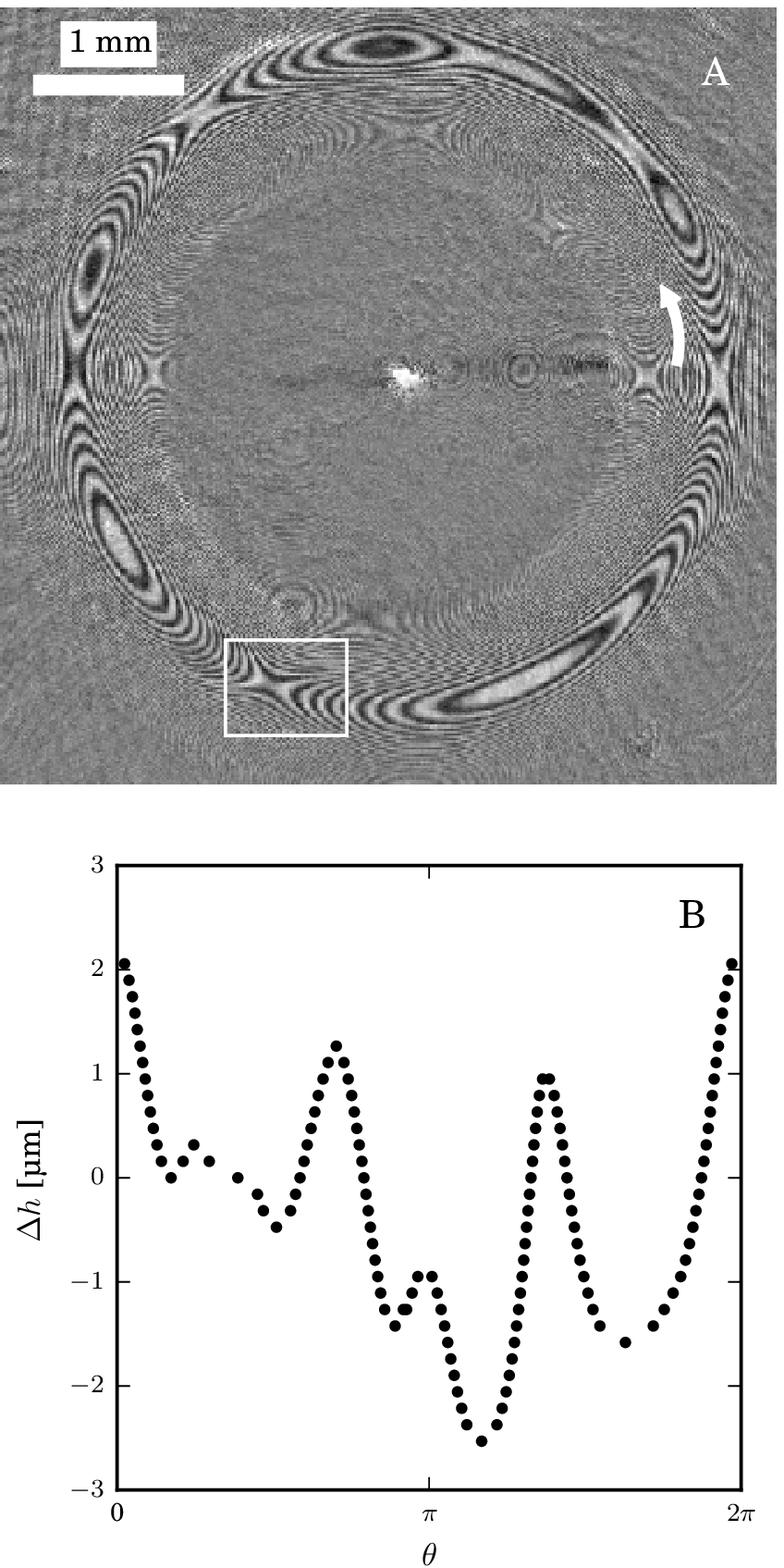}
  \caption{A) A typical normalized interference image.  The region of interest is around the rim, where the drop is closest to and almost parallel to the surface.  The patterns inside and outside of the main rim are Moiré patterns and should be ignored.  The peak features on the rim are the regions where the rim is closest to the surface and the saddles are the regions where the rim is farthest from the surface.  The height difference between adjacent dark and light fringes is $|\lambda_{l}/4|$.  By counting fringes, one can determine the height profile around the rim. B) The height profile extracted for this frame.  The arrow in A indcates the location of $\theta = 0$ and the direction of winding.  The box indicates the region shown in Fig.~\ref{fig:fringe_find}.}
  \label{fig:raw_unmarked}
\end{figure}

A typical normalized interference image is shown in Fig.~\ref{fig:raw_unmarked}A.  The largest, clearest fringes in the image correspond to the rim region marked in Fig.~\ref{fig:drop_scheme}, where the bottom of the drop is most nearly parallel to the top of the prism.  As the surface becomes steeper, both inside and outside of the rim, the fringes become more tightly spaced.  This results in Moiré patterns due to interference between the fringe pattern and the pixels of the camera when their spatial frequencies become comparable.  Because the drop surface must be continuous,  eq.~(\ref{eq:interference}) implies that contiguous regions of equal intensity are surfaces of constant $h$.  Further, adjacent pairs of light and dark fringes, where the intensity has moved from a local maximum to a local minimum, have a height difference of $\left|\lambda_{l} / 4\right|$.

In general, interference only gives the absolute value of the height difference between adjacent fringes, not the sign of the difference.  However in this case the curvature in the radial direction is always upward, as shown in Fig.~\ref{fig:drop_scheme}.  Thus the peak features, where the Gaussian curvature is positive, are where the rim is closest to the surface and the saddle features, where the Gaussian curvature is negative, are where the rim is farthest from the surface.  Further, because the surface of the drop is continuous, all of the steps between a saddle and a peak are in the same direction.  Thus, by starting on any fringe and counting fringes around the rim one can build up the profile of the rim, as was done by hand in~\cite{burton12:_geomet_vapor_layer_under_leiden_drop}.  Here a method is presented to automate this process.

An example profile extracted by my software is shown in Fig.~\ref{fig:raw_unmarked}B.   $\theta$ is the angle around the rim as measured from the x-axis and the direction of the winding is marked on Fig.~\ref{fig:raw_unmarked}A with the arrow.  As one goes around the rim one can follow the profile, including the two-fringe high local maximum at $\theta \approx \pi / 4$.  Using this technique one can, in principle, resolve variations in the height of the gap under the the rim to $\lambda_{l}/4 = 158\textrm{nm}$.

\subsection{Image Analysis}
This paper extends the work of Burton et al~\cite{burton12:_geomet_vapor_layer_under_leiden_drop} by automating the image analysis to identify the interference fringes and reconstruct the rim shape and profile from high-speed interference-image movies.  This allows a measurement of the rim size, shape, location and height profile as a function of time.  The following sections will describe how to locate and classify fringes (\ref{subsec:FL}) in each frame of a movie.  The fringe locations are first used to reconstruct the rim shape at a given time and then used to reconstruct the height profile (\ref{subsec:HR}) as a function of time.  The code used for this analysis is available under the GPL at \url{https://github.com/tacaswell/leidenfrost}.

\subsubsection{Fringe Location}
\label{subsec:FL}
\begin{figure*}
  \includegraphics{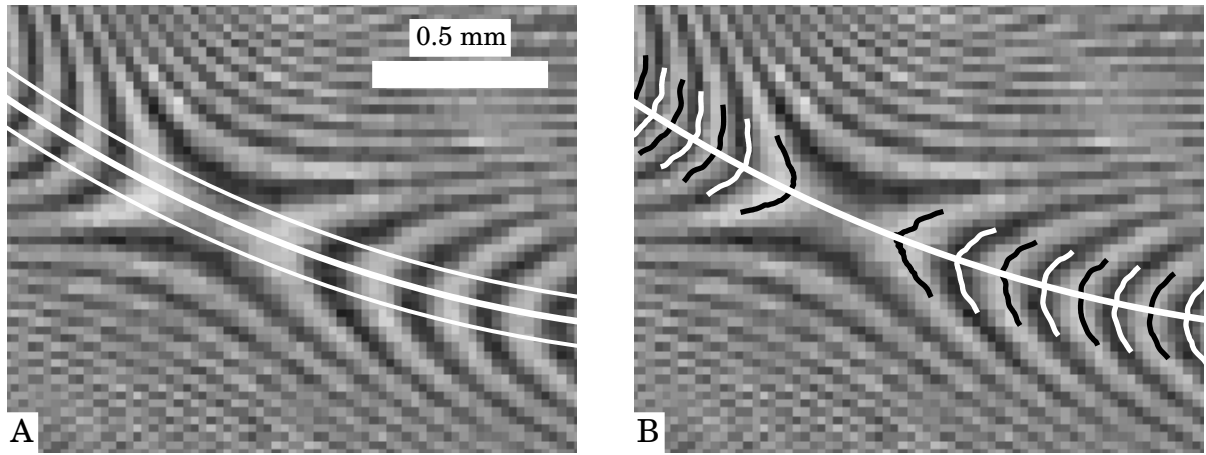}
  \caption{Detail of the area of the rim marked with a box in Figure.\ \ref{fig:raw_unmarked}. A) Image shows the initial seed-curve as a thick white line along the rim.  Two scaled-curves are shown as thinner white lines.  B) Image shows the traces of the located fringes as white lines on the dark fringes and black lines on the light fringes.  The solid white curve along the rim is the rim shape as determined from the fringes.}
  \label{fig:fringe_find}
\end{figure*}

To locate the fringes, the algorithm needs to be provided with a seed-curve which approximately traces the rim in the image.  For the first frame this is supplied by hand, but for all subsequent frames the result from the previous frame is used. This allows the code to process long time series with no supervision.  The seed-curve is represented with a periodic one-parameter spline, with parameter $\xi$.  The seed-curve, $\kappa_{0}(\xi)$, is used to generate a family of scaled-curves paramterized by $s$ with $\kappa_{s}(\xi) = \kappa_{0}(\xi) + s \hat{n}(\xi)$ where $\hat{n}(\xi)$ is the outward pointing unit-normal to $\kappa_{0}$ at $\xi$.  This process is analogous to generating a ruled surface.  The seed-curve, $\kappa_{0}$ (thicker white line), and a pair of scaled-curves, $\kappa_{2}$ and $\kappa_{-2}$ (thinner white lines), are shown in Fig.~\ref{fig:fringe_find}A.  The region of the rim shown is marked with a box in Fig.~\ref{fig:raw_unmarked}A.  The interference image is then sampled along each curve, $\kappa_{s}$, in the family and the local maximum and minimum of the intensity along the slice are identified as a function of $\xi$.  The locations of the maximum and minimum are connected between curves using a variation on the Crocker-Gorier algorithm~\cite{crocker95_meth_dig} commonly used in single particle-tracking~\footnote{For a pure-python implementation of Crocker-Grier see \url{https://github.com/soft-matter/trackpy}}.  The resulting tracks trace the chevron shaped fringes, which are the features of interest.  Examples of the identified fringes are shown in Fig.~\ref{fig:fringe_find}B with black curves on the light fringes and vice versa.  The fringes are then classified by color and by which direction around the rim they `point'.  The rim is located by generating a spline using the `tips' of the fringes.  The generated spline is shown in Fig.~\ref{fig:fringe_find}B as the thick white line.  This spline is used as the seed-curve for the next frame and the process repeats.  Using the spline representation of the rim shape one can convert the coordinates of a point on the rim between the paramaterized coordinates, $(\xi, s)$, Cartesian image coordinates, $(x, y)$, and Polar coordinates, $(r, \theta)$ where $r$ is the distance from the rim to the center of the drop and $\theta$ is the angle the line from the center to the rim makes with respect to the $x$-axis.

\subsubsection{Height Reconstruction}
\label{subsec:HR}
\begin{figure}
  \includegraphics{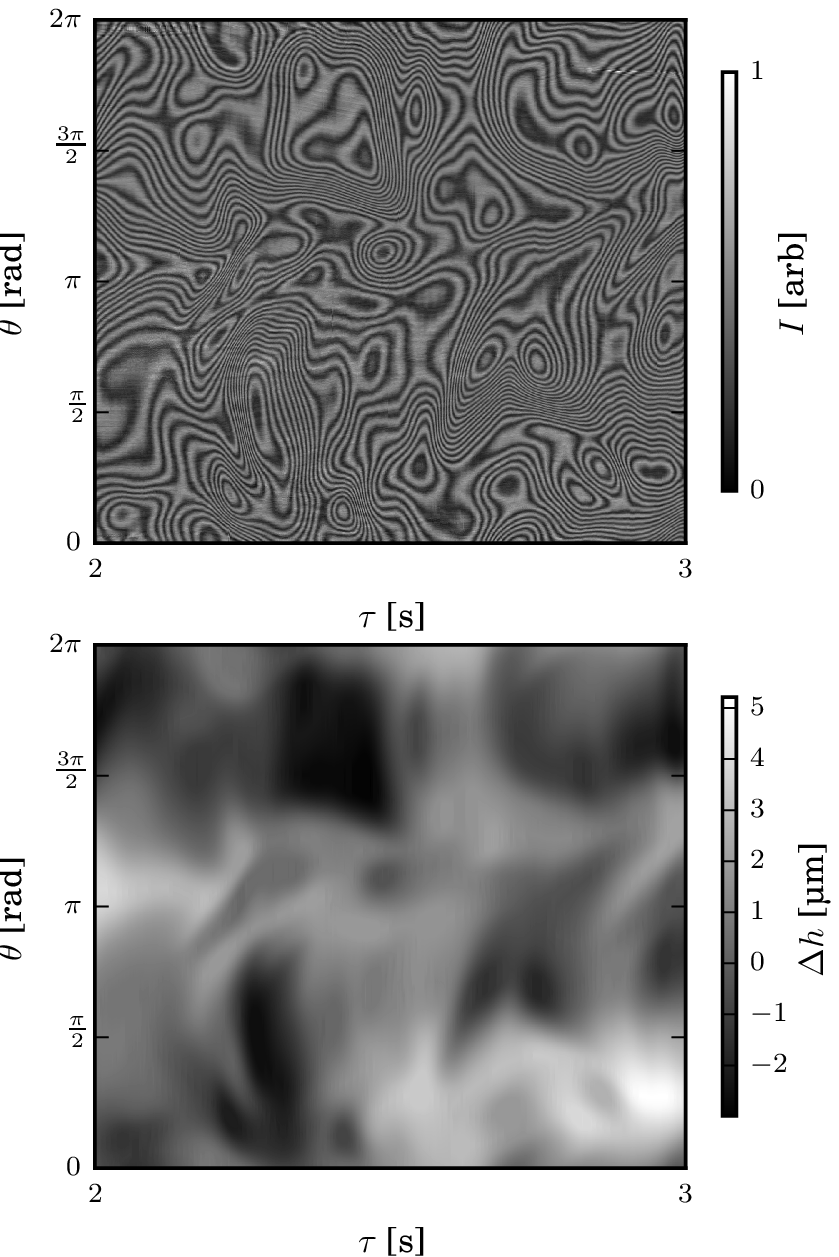}
  \caption{A)  Kymograph of the raw data.  Each column of this image is generate by sampling the interference images along the rim, as identified by the fringes.  Contiguous regions are surfaces of constant height.  The fringe data determines relative height between adjacent regions. B) Kymograph of $\Delta h(\theta, \tau)$ for the data in A.  Some features can easily be matched between the panels, such as the large basin at $(3/2 \pi, 2.4)$. }
  \label{fig:raw_kymo}
\end{figure}

To reconstruct the height profile of the rim through time, it is best to incorporate information from many frames.  This is done by generating a kymograph, which is a generic two dimensional space-time plot. A typical example is shown in Fig.~\ref{fig:raw_kymo}A.  In this case, each column is a slice through an interference image along the rim identified from the fringes.  The vertical axis is the angle around the rim, $\theta$, and the horizontal axis is time, $\tau$.  Moving vertically in the kymograph moves around the rim of the drop at a fixed time and moving horizontally in the kymograph stays at a fixed $\theta$ on the rim, but moves between frames.  As with the interference image shown in Fig.~\ref{fig:raw_unmarked}, contiguous regions of the same intensity are at the same height above the surface and adjacent regions of the alternate color have a height difference of $\left|\lambda_{l} / 4\right|$.  Standard image processing tools  can segment the kymograph into light and dark regions and the relative height difference between adjacent regions is determined using the information from the fringes.  Starting from an arbitrary region, which is assigned $\Delta h \equiv 0$, a $\Delta h$ value is assigned iteratively to each region.  This reconstructs the height profile, $\Delta h(\theta, \tau)$, relative to a fixed reference height as a function of angle around the rim, $\theta$, and time, $\tau$.  A kymograph of $\Delta h(\theta, \tau)$ is shown in Fig.~\ref{fig:raw_kymo}B for the data shown in Fig.~\ref{fig:raw_kymo}A.  Although this method cannot measure the absolute height from the surface to the drop, I am able to track how every point on the rim  moves toward or away from the surface as a function of time.

\section{Results}
\label{sec:results}

\subsection{Rim Shape Vibrations}
\label{sec:rim-shape-vibrations}
\begin{figure}
  \centering
  \includegraphics{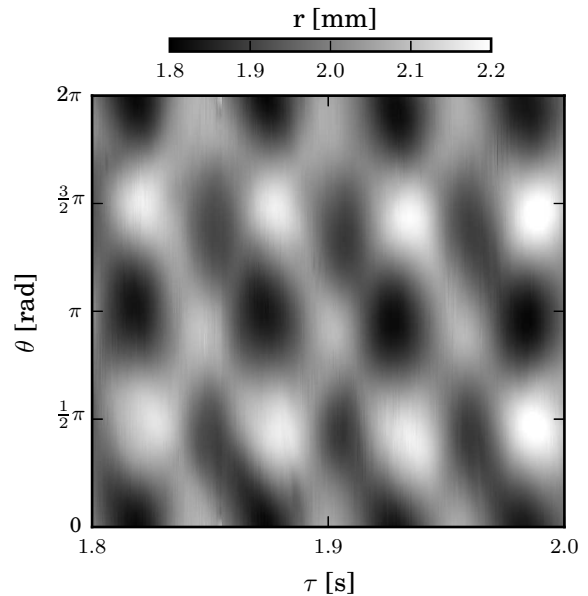}
  \caption{Plot of $r(\theta, \tau)$ for a drop showing a strong $n=2$ oscillation.  The vertical axis is $\theta$, the angle around the rim, the horizontal axis $\tau$ is time and the color indicates the radius $r(\theta, \tau)$.  The standing $n=2$ oscillations is clearly visible as the square pattern of light and dark regions.}
\label{fig:shape_kymo}
\end{figure}

\begin{figure}
  \centering
  \includegraphics{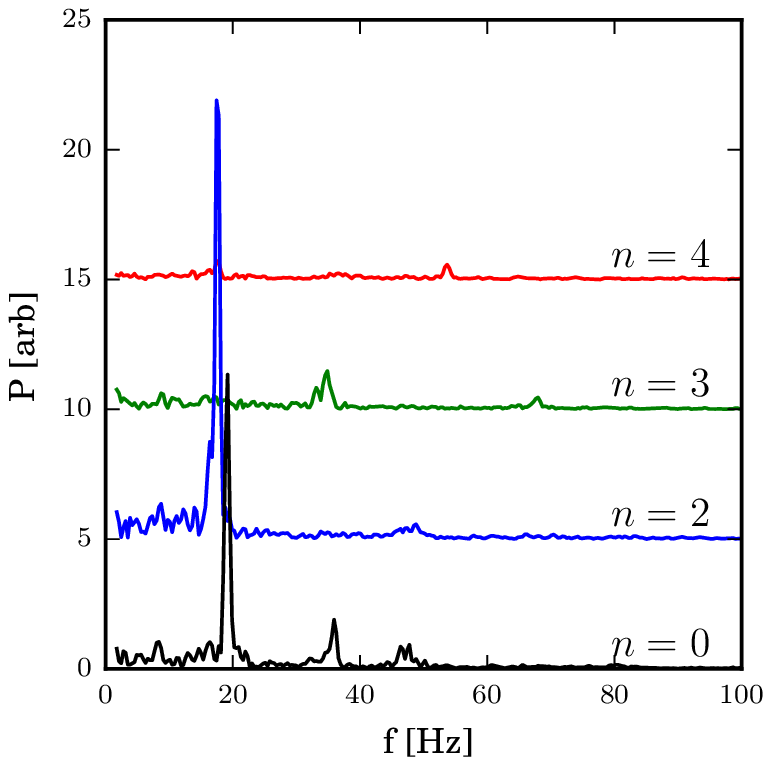}
  \caption{(Color online) Fourier spectrum of the rim shape for Fourier modes $n = \{0, 2,3,4\}$ for the drop shown in Fig.~\ref{fig:shape_kymo}.  The curves have been shifted vertically for clarity.  There are strong peaks in $n=0$ and $n=2$, and smaller, but clear, peaks in modes $n=3$ and $n=4$.}
  \label{fig:rimshape_fft}
\end{figure}
The existence of large amplitude vibrations in the horizontal shape of Leidenfrost drops was first reported in the 50's~\cite{holter52:_vibrat_evapor_liquid_drops} and has been thoroughly studied in a variety of systems since then~\cite{snezhko08:_pulsat,Strier2000261,bouwhuis13:_oscil,tokugawa94:_mechan_self_induc_vibrat_liquid,takaki85:_vibrat_flatt_drop,adachi84:_vibrat_flatt_drop}.  For drops with $R > \lambda_{c}$ the vibrational spectrum can be approximated by the vibrational modes of a cylinder of fluid, which were predicted by Lord Rayleigh~\cite{Rayleigh01011879} to be
\begin{equation}
f_{n} = \frac{1}{2\pi}\sqrt{\frac{\gamma n (n^{2} - 1)}{\rho R^{3} } }.
\label{eq:fn_pancake}
\end{equation}
where $n$ is the number of wavelengths around the drop circumference.  For small drops, the vibrational spectrum is given by a slightly different formula for waves on a sphere~\cite{Rayleigh01011879}
\begin{equation}
f_{n} = \frac{1}{2\pi}\sqrt{\frac{\gamma n  (n -1) (n + 2)}{\rho R^{3} } }
\label{eq:fn_marble}
\end{equation}

The shape of the rim as a function of angle and time is given by $r(\theta, \tau)$.   A kymograph of $r(\theta, \tau)$ for a drop with a $n=2$ vibration is shown in Fig.~\ref{fig:shape_kymo}.  As with Fig~\ref{fig:raw_kymo} the axis are $\theta$ and $\tau$, however in this case the intensity represents the distance from the center to the rim.  Along the vertical direction the radius has two local maximum and two local minimum indicating that this is a $n=2$ deformation.    The checker-board pattern, with local maximum turning in to local minimum as a function of $\tau$, is characteristic of a standing wave.  An alternate way to understand the pattern in Fig.~\ref{fig:shape_kymo} is to note the two sets of diagonal lines, one up to right and one down to the right.   These indicate the drop has two counter-propagating traveling waves,, which is exactly the mathematical description of a standing wave.

To quantify the rim shape dynamics, one can take the Fourier transform of $r(\theta, \tau)$,
\begin{equation}
  \label{eq:r_fft}
  \hat{r}(n, f) = \iint\,d\theta d\tau\, r(\theta, \tau) e^{\imath \theta n} e^{\imath 2\pi f \tau}
\end{equation}
where $n$ is the spatial mode number and $f$ is the frequency.  It is important to note that the spatial transform is between $n$ and $\theta$, not between the wavenumber $k$ and distance along the rim.  The rim is periodic, thus for a given circumference, $c$, only discrete wavenumbers, $k_{n}=2\pi n / c$, are allowed.  However the circumference of the rim is varying in time so that it is difficult to consider the Fourier transform at fixed $k$.  The strong peak in the $n=2$ spectrum is the oscillation clearly visible in Fig.~\ref{fig:shape_kymo}.  In addition to the peak in  $n=2$ oscillation there are also small but distinct peaks in $n=3$ and $n=4$ and a large peak in $n=0$.  There is no $n=1$ spectrum because $n=1$ corresponds to a shift in the location of the center of the rim which is not captured using this description of the rim shape.

\begin{figure}
  \centering
  \includegraphics{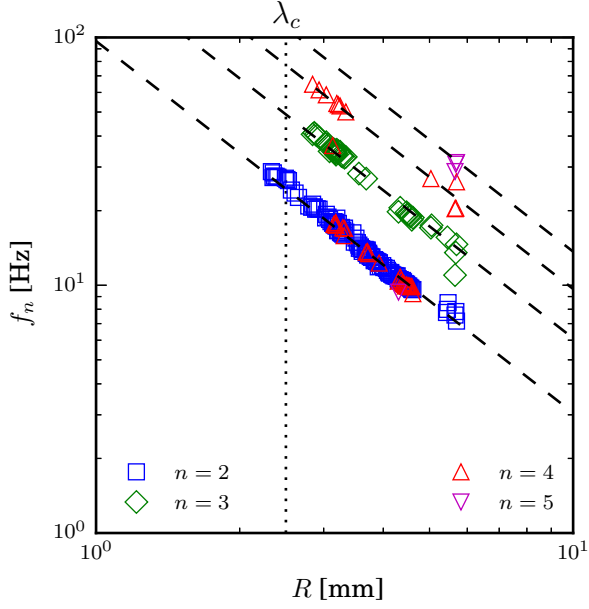}
  \caption{(color online) The vibrational modes versus drop radius.  The  frequencies scale with drop radius and mode number as expected.  The points are the location of the maximum in $\hat{r}(n, f)$ for $n=2$ (squares), $n=3$ (diamonds), $n=4$ (up triangles), and $n=5$ (down triangles) as a function of $R$.  The black dashed lines are the zero-parameter predictions from eq.~(\ref{eq:fn_pancake}) for $n = \{2, 3, 4, 5\}$ from the bottom to the top.  The vertical dashed line indicates $R = \lambda_{c}$.}
  \label{fig:vibration_v_radius}
\end{figure}

I preformed this analysis on a range of drop sizes and identified the largest peak in $\hat{r}(n, f)$, $f_{n}$, for each Fourier modes $n = \{2, 3, 4, 5\}$.  These frequencies are plotted in Fig.~\ref{fig:vibration_v_radius} versus $R$, the Fourier mode is indicated by the marker shape and color.  The frequencies predicted by eq.~(\ref{eq:fn_pancake}) are plotted as the dashed lines for $n = \{2, 3, 4, 5\}$, from the bottom to top, with no fitting parameters.  There is excellent agreement between these measurements and those predicted and previously measured.  The data shown spans a range of surface temperatures and over two orders of magnitude in oscillation amplitude.  At small amplitudes the vibrations are linear and well described by pure sine-waves.  The modes are uncoupled and fall on the line predicted by eq.~(\ref{eq:fn_pancake}).  At large amplitudes, the vibrations become non-linear and the drop shape is no longer described by pure sine waves which induces coupling between the modes, as demonstrated by the $n=4$ symbols that fall on to the $n=2$ and $n=3$ lines.  In drops with large amplitude oscillations it is also common to see multiple peaks in the spectrum of the modes, as in Fig.~\ref{fig:rimshape_fft}.

The vertical line in Fig.~\ref{fig:vibration_v_radius} shows where $R = \lambda_{c}$.  This is where the drop changes shape regime between a dented sphere and a cylinder.  Although I observed drops with $R < \lambda_{c}$, none of those have detectable shape oscillations in mode $n=2$ or higher.

\subsection{Breathing mode}

\begin{figure}
  \centering
  \includegraphics{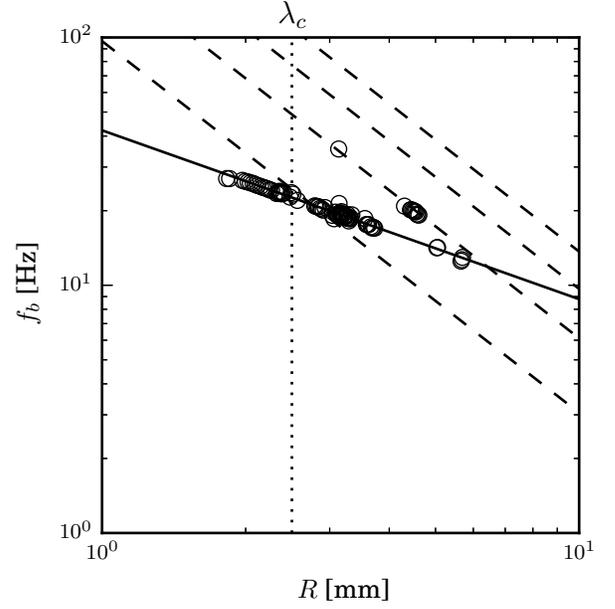}
  \caption{The frequency of the breathing mode, $f_{b}$ as a function $R$.  The dashed lines are the predictions from eq.~(\ref{eq:fn_pancake}) for $n=\{ 2, 3, 4, 5\}$.  In drops with large amplitude vibrations the breathing mode can become coupled with higher modes as shown by the points falling on the $n=3$ line.  The solid line is a power law fit with $f_{b} \propto R^{-0.68\pm 0.01}$.}
\label{fig:breathing_v_radius}
\end{figure}

In addition to the predicted vibrational modes, a breathing mode is also observed in the average rim radius, shown as $n=0$ mode in Fig.~\ref{fig:rimshape_fft}.  However, such a mode is not predicted by eqs.~(\ref{eq:fn_pancake})~and~(\ref{eq:fn_marble}).  This breathing mode is a robust feature of Leidenfrost drops; I observe it at some magnitude in almost every drop studied.  The largest peak in $\hat{r}(0, f)$, $f_{b}$, is plotted versus $R$ in Fig.~\ref{fig:breathing_v_radius}.  As noted above, large amplitude vibrations can couple modes together.  This is the case for the points that fall on the $n=3$ line.   Excluding those points, I can fit a power law:
\begin{equation}
  \label{eq:fb_powerlaw}
  f_{b} \propto R^{-0.68\pm 0.01},
\end{equation}
shown as the solid line in Fig.~\ref{fig:breathing_v_radius}. The scaling is very different than the $R^{-3/2}$ scaling in eq.~(\ref{eq:fn_pancake}) and eq.~(\ref{eq:fn_marble}) and does not appear to depend on whether $R$ is above or below $\lambda_{c}$.

\begin{figure}
  \includegraphics{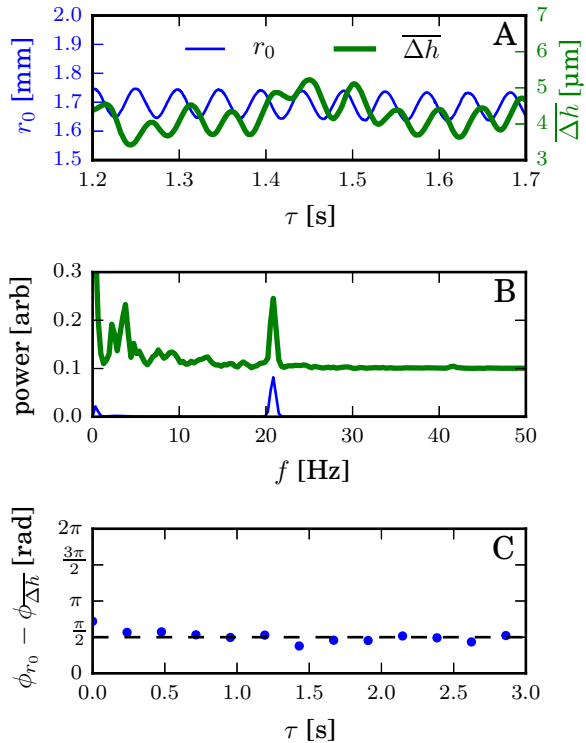}
  \caption{(color online) Coupling of the drop radius and veritcal height of the drop.  A) A 0.5s time trace of the rim circumference, $c$, is shown in blue on the left axes with a thin line and $\overline{\Delta h}$ is shown in red on the right axes with a thick line. B) The power spectrum of the curves in A) are shown with the same colors.  The spectrum for the height is shifted vertically by $0.1$ for clarity.  C) The phase difference between the curves in A) are shown as a function of time.  The signals are phase-locked with a difference of $\pi/2$ for the full length of the movie.}
  \label{fig:phase_lock}
\end{figure}

In some drops there is a clear coupling between the average rim radius, $r_0(\tau)$, and the mean height, $\overline{\Delta h}(\tau) = \frac{1}{2\pi} \int d\theta\, \Delta h(\tau, \theta)$, around the rim.    In Fig.~\ref{fig:phase_lock}A I show a 0.5s trace of $r_{0}(\tau)$ (on the left axis and the thin blue line) and $\overline{\Delta h}(\tau)$ (on the right axis and the thick green line) which appear to be phase locked to each other.  Figures~\ref{fig:phase_lock}B show the Fourier transforms of both curves; both have a sharp peak at 21Hz.  The spectrum for the height is shifted vertically by $0.1$ for clarity.  This high-frequency oscillation is on top of lower-frequency oscillations.  To quantify the phase-locking between the curves I compute the Fourier coefficient over windows of $5$ periods at $f_{b}$:
\begin{equation}
  \label{eq:ff_component}
  A_{\chi} = \int_{\tau_{0}}^{\tau_{0} + 5 / 2\pi f_{b}} d\tau\, e^{\imath\tau2\pi f_{b}} \chi(\tau) = |A|e^{\phi_{\chi}(\tau_{0})}
\end{equation}
where $\chi$ is either $r_0$ or $\overline{\Delta h}$.  In Fig.\
\ref{fig:phase_lock}C I plot $\phi_{r_{0}} -
\phi_{\overline{\Delta h}}$, which shows that $r_0(\tau)$
consistently leads $\overline{\Delta h}(\tau)$ by $\pi/2$ over a 3s time scale, the full length of the movie.

To understand the relative phase between $r_{0}(\tau)$ and $\overline{\Delta h}(\tau)$, I consider a simple one dimensional model.   The model assumes the vertical motion of the drop is in the over-damped regime, due to the resistance to motion being dominated by the vapor escaping under the rim where $\rm{Re} \ll 1$, and that the pressure in the vapor region is approximately constant over the course of a cycle.  The only forces acting on the drop are pressure pushing upward and gravity pulling down.  With these assumptions:
\begin{equation}
  \label{eq:free_od}
  F_{\rm{net}} = \pi r_{0}^{2} p - mg \propto \frac{d \overline{\Delta h}}{d t}
\end{equation}
where $p$ is the pressure in the vapor layer and $m$ is the mass of the drop.  Assuming:
\begin{equation}
  \label{eq:r_ansatz}
  r_{0} = \overline{r_{0}} + \alpha \sin\left(2\pi f t +  \phi_{r_{0}}\right)
\end{equation}
 and
 \begin{equation}
   \label{eq:h_ansatz}
 \overline{\Delta h} = \beta \sin\left(2\pi f t +  \phi_{\overline{\Delta h}}\right)
 \end{equation}
 where $\alpha$ and $\beta$ are the amplitudes of the respective variations.  $\overline{r_{0}} = \sqrt{mg/p\pi}$, such that when $\alpha = 0$, $F_{\rm{net}} = \beta = 0$.  Plugging eqs.~(\ref{eq:r_ansatz}) and~(\ref{eq:h_ansatz}) into eq.~(\ref{eq:free_od}) and simplifying to leading order in $\alpha$ gives
\begin{equation}
  \label{eq:free_od_simplified}
  \frac{\overline{r_{0}} \alpha}{f \pi} \sin\left(2\pi f t + \phi_{r_{0}}\right) \propto
\beta \sin\left(2\pi f t +  \phi_{\overline{\Delta h}} + \frac{\pi}{2}\right)
\end{equation}
which implies that
\begin{equation}
  \label{eq:phi_prediction}
  \phi_{r_{0}} - \phi_{\overline{\Delta h}} = \frac{\pi}{2}
\end{equation}
as observed.  Although this simple model matches the experimental results, it does not take into account any effects related to flow within the drop, nor account for any time dependence of the pressure, the volume of the vapor layer, or the flux into and out of the vapor layer.  Simulations~\cite{bouwhuis13:_oscil} have shown that the drop can undergo an axi-symmetric oscillation coupled to vertical motion, but have not observed a phase shift.

\subsection{Qualitative changes to rim profile}

In addition to tracking the rim's average height, I also studied its profile as a function of time.  Qualitatively, large drops have `active' height profiles  with multiple local minima and maxima around the rim that evolve in time.  In contrast small drops tend to have a dominant local maximum-minimum pair which does not significantly fluctuate in location or amplitude.  Given how well capillary waves describe the in-plane vibrations of the rim shape it is natural to compare the height fluctuations of the rim to capillary-wave motion  as well.  However, the typical time scale for the rim height profile to evolve, even in large drops, is much slower than the period of a capillary-wave with a wavelength on the scale of the rim circumference, 8-15 mm.  Unlike the rim shape dynamics, discussed in sec.~\ref{sec:rim-shape-vibrations}, the rim height profile does not show clear standing- or traveling-wave patterns.

\begin{figure}
  \includegraphics{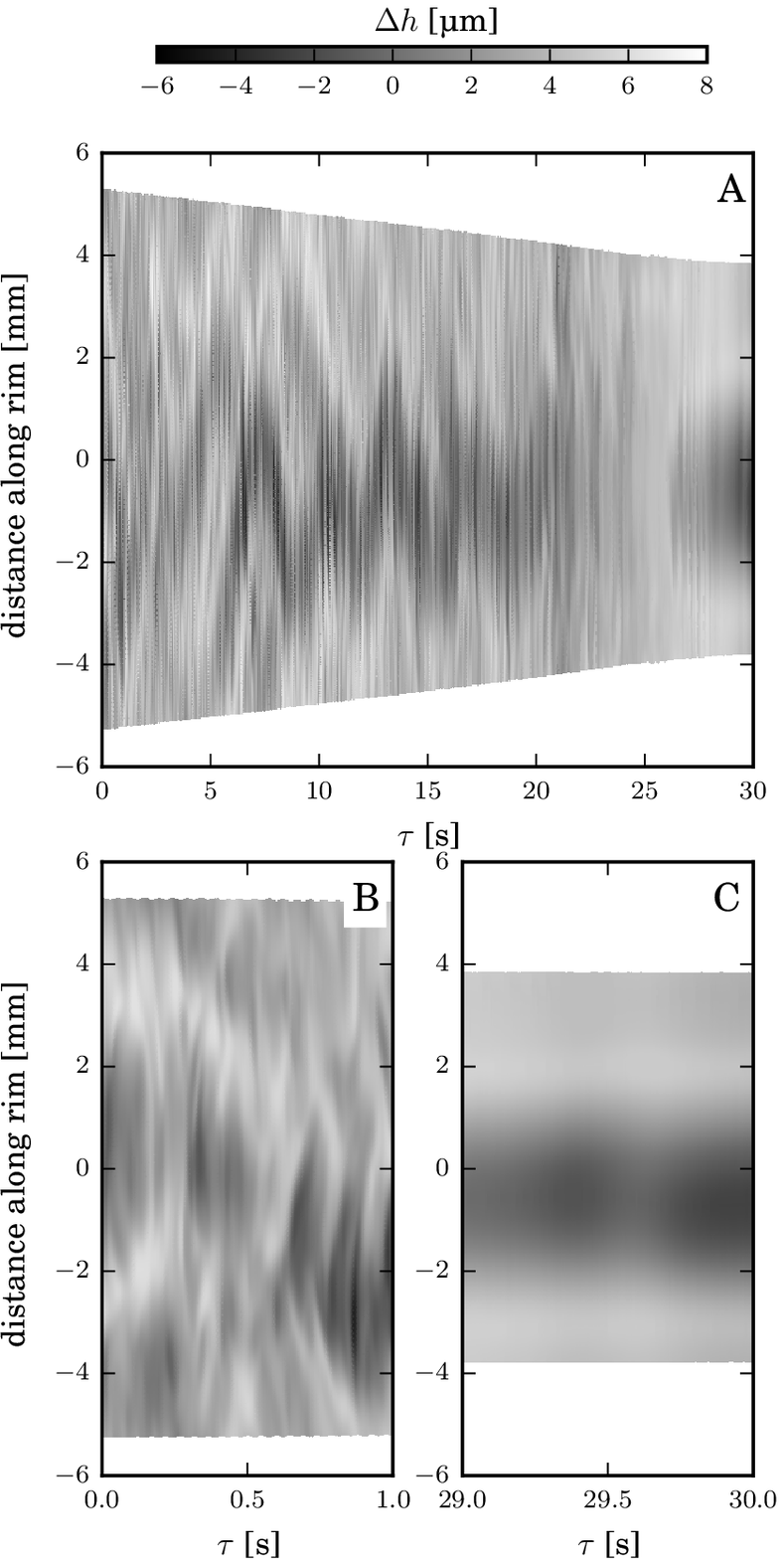}
  \caption{Visualization of evolution of the height height profile.  The vertical axis is distance along the rim, the horizontal axis is time.  A) Shows $30$s of data where the secular decrease in the circumference due to evaporation is clearly visible.  B) and C) show the details of first and last $1.0$s intervals respectively.  Comparing B and C, it is clear that there is a qualitative change in the structure and dynamics of the height profile as a function of drop size.}
  \label{fig:texture}
\end{figure}

By taking long movies (over 30 seconds in duration) I was able to capture the transition from the large active regime to the small quiescent regime in a single drop.  Figure~\ref{fig:texture}A  shows the full evolution of the height profile as a function of time.  Unlike the kymographs shown in Fig.~\ref{fig:raw_kymo}~and~\ref{fig:shape_kymo} where $\theta$ the vertical axis,  the panels in Fig.~\ref{fig:texture} use distance along the rim for the vertical axis.  Thus, as the circumference of rim varies in time the width of the kymograph `ribbon' varies.  The large scale secular decrease in the width in Figure~\ref{fig:texture}A is due to drop shrinking from evaporation.

To better see the short-time structure and dynamics of the height profile details of the first and last $1.0$s intervals of Figure~\ref{fig:texture}A are shown in Figure~\ref{fig:texture}B and C respectively.  The active dynamics of the large drop can be clearly seen in the texture of Figure~\ref{fig:texture}B, which is qualitatively similar to Figure~\ref{fig:raw_kymo}B.  Unlike Fig.~\ref{fig:shape_kymo}, there are no checker-board pattern or diagonal strips indicating that unlike the rim shape, the rim height profile is not dominated by standing or traveling waves.  The small ripples along the top and bottom edges of the kymograph in Figure~\ref{fig:texture}B are the due to the breathing mode discussed above.  Is contrast the height profile in C has a single pair of local maximum/minimum which do not significantly fluctuate over the course of a second.

\begin{figure}
  \centering
  \includegraphics{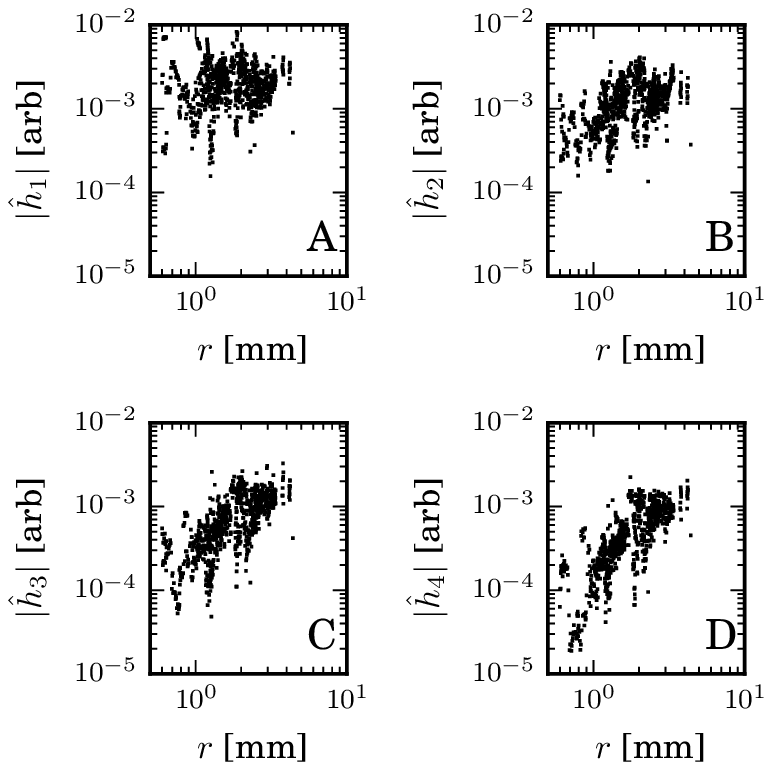}
  \caption{Average power in the first four Fourier modes in the rim height profile over .5s windows for a range of drop sizes.  The power in $j=1$ is independent of drop size, where as the higher modes are suppressed as $r$ decreases. }
  \label{fig:ruffle_fft}
\end{figure}

To quantify the change in the structure as a function drop size, I compute the Fourier components:
\begin{equation}
  \label{eq:profile_fft}
  \hat{h}_{j}(\tau) = \int_{0}^{2\pi}d\theta\,\Delta h(\theta, \tau)e^{\imath j \theta}
\end{equation}
where $j$ is an integer.  As above, the components are computed between $j$ and $\theta$, instead of length and wavenumber because the allowed wavelengths vary in time.  In Figure~\ref{fig:ruffle_fft} the average of $|\hat{h}_{j}|$ over $0.5$s windows is plotted against the rim radius, $r$, for the first four Fourier modes, $j=\{1, 2. 3, 4\}$.   The $j=1$ mode is a tilt in the bottom surface of the drop, relative to the hot surface.  Because tilting the surface does not create any additional curvature around the rim, the energy cost to at a fixed amplitude is independent of the rim size.   This is in agreement with Figure~\ref{fig:ruffle_fft}A which shows that  $|\hat{h}_{1}|$ is independent of $r$.  In contrast,  $n\geq 2$ require additional curvature because they introduce bending on the rim in the azimuthal direction.  As the drop evaporates and the wavelength at a fixed $j$ decreases, thus increasing the amount of curvature, and hence energy, required for a given amplitude.  This is in agreement with Figures~\ref{fig:ruffle_fft}B, \ref{fig:ruffle_fft}C, and~\ref{fig:ruffle_fft}D, which show $|\hat{h}_{2}|$, $|\hat{h}_{3}|$ and $|\hat{h}_{4}|$ decaying with decreasing drop size.

I never observed a drop that stable while flat and parallel to the prism.  There was always some degree of tilt or height profile fluctuation around the rim.  A possible explanation for this observation is that the drop is able to lower it's center of mass by introducing an asymmetry in the rim height profile.  The flux through a very narrow gap goes as the cube of the gap height, $Q_{\rm{out}} \propto h^{3}$, thus if two sections of the rim are raised and lowered by the same small amount, to not change the center of mass of the drop, there will be a net increase in the flux out from the high-pressure region.  To maintain the balance between flux into and out of the vapor region the drop will reduce the height of it's center of mass, which will both increase the evaporative flux in and decrease the flux out, and lower the potential energy of the whole system.  Thus the axisymmetric case is unstable to small perturbations in the rim height profile and the drop will always be have some height variation around the rim.  This instability is presumably cut off by the details of the vapor flow and surface tension which prevent any part of the drop from getting too close to the surface and prevent large curvatures from developing along the rim.

\section{Conclusions}

This paper has extended the high-speed laser-interference technique to study the bottom surface of a Leidenfrost drop.  By automating the image analysis to locate, classify, and interpret the interference fringes it has been possible to extract the location, shape, and height profile of the rim in each frame of a high-speed movie.  From this, the frequencies of oscillation for the shape of the rim were obtained.  Moreover, this capability has allowed the observation of a number of previously unreported dynamics such as a breathing mode in the size of the rim which is coupled to, and $\pi/2$ out of phase with, the vertical motion of the drop.

The rim, where the drop is closest to the surface, is never a uniform height above the substrate but is unstable against small perturbations so that there will always be a variation in the gap height around the rim. The absolute magniture of the height fluctuation does not appear to depend on drop size. This points to an underlying instability driven by the increased ability of gas to escape the pocket when the rim is non-uniform. Whereas the in-plane shape deformations are associated with capillary waves, the variations in the gap height are not. The timescale on which the gap heights fluctuate is an order of magnitude slower than capillary waves of comparable wavelength and there is no evidence of persistent standing or traveling waves around the rim. The dynamics of the gap height are controlled by a combination of the dynamics of the thin-film vapor flow of the escaping gas and the dynamics associated with the overall drop shape. Large drops are `active' with multiple pairs of local extrema that evolve in both location and size as a function of time; small drops have a single dominant pair of extrema that are relatively fixed in position and size. The onset of the quiescent small-drop regime is surprizingly sudden. This clearly shows that the Leidenfrost vapor layer has complex and rich dynamics that warrants further investigation.

\section{Acknowledgments}
I am grateful to Sidney R Nagel for helpful discussions and advice.  I am also grateful to Justin Burton who's experimental apparatus I adapted for this work.  This work was supported by NSF Grant DMR-1105145 and NSF-MRSEC DMR-0820054.

\bibliographystyle{apsrev4-1}

\end{document}